\documentclass[journal]{rmaa}

\newcommand{\mysp}{}\def\mysp/{}
\newcommand{\teff}{}\def\teff/{$T_{\mathrm{eff}}$\xspace}
\newcommand{\ha}{H$\alpha$\xspace}
\newcommand{\hb}{H$\beta$\xspace}
\newcommand{\hii}{\ion{H}{ii}\xspace}
\newcommand{\he}{\ion{He}{i}\xspace}
\newcommand{\sii}{[\ion{S}{ii}]\xspace}
\newcommand{\siip}{[\ion{S}{ii}]+\xspace}
\newcommand{\nii}{[\ion{N}{II}]\xspace}
\newcommand{\oi}{[\ion{O}{I}]\xspace}
\newcommand{\oii}{[\ion{O}{II}]\xspace}
\newcommand{\oiii}{[\ion{O}{III}]\xspace}
\newcommand{\lhe}{\ion{He}{I}$\lambda5876$\xspace}
\newcommand{\lsii}{[\ion{S}{II}]$\lambda6716$\xspace}
\newcommand{\lsiimas}{[\ion{S}{II}]$\lambda6716+\lambda6731$\xspace}
\newcommand{\lnii}{[\ion{N}{II}]$\lambda6583$\xspace}
\newcommand{\loi}{[\ion{O}{I}]$\lambda6300$\xspace}
\newcommand{\loii}{[\ion{O}{II}]$\lambda3727$\xspace}
\newcommand{\loiii}{[\ion{O}{III}]$\lambda5007$\xspace}
\newcommand{\ddb}{{\sc DIGEDA}\xspace}

\usepackage{xspace}

\title{Analysis of the diffuse ionized gas database: DIGEDA}

\author{
Nahiely Flores-Fajardo, Christophe Morisset, and Luc Binette\\
Instituto de Astronom\'ia, Universidad Nacional Aut\'onoma de M\'exico, Mexico}

\shortauthor{Flores-Fajardo, Morisset, \& Binette}

\shorttitle{\ddb: a DIG observational database}

\listofauthors{N.~Flores-Fajardo, C.~Morisset, \& L.~Binette}

\indexauthor{Flores-Fajardo, N.}
\indexauthor{Morisset, C.}
\indexauthor{Binette, L.}

\fulladdresses{
\item Luc Binette, Nahiely Flores-Fajardo, and Christophe Morisset: Instituto de Astronom\'ia, 
Universidad Nacional Aut\'onoma de M\'exico, Apdo. Postal 70-264 04510, M\'exico D.F., Mexico 
(nahiely,~morisset@astroscu.unam.mx).}

\ReceivedDate{2009 June 17} 
\AcceptedDate{2009 August 15} 
\SetYear{2009}

\resumen{Los estudios del Gas Ionizado Difuso (DIG), hasta el momento se han hecho 
sin un consenso del criterio estricto para diferenciar entre el DIG y las regiones 
\hii. En este trabajo recopilamos las mediciones de las l\'{i}neas en emisi\'on de 
29 galaxias disponibles en la literatura, creando la primera base de datos del DIG 
(\ddb). Haciendo uso de esta base, analizamos las propiedades globales del DIG a 
partir de los cocientes de l\'{i}neas \lnii/\ha, \loi/\ha, \loiii/\hb y \lsii/\ha, 
as\'{i} como la medida de emisi\'on de \ha. Este an\'alisis nos permiti\'o concluir 
que el cociente \nii/\ha es un criterio general para diferenciar si una regi\'on en 
emisi\'on es DIG o regi\'on \hii, mientras que la EM(\ha) es una cantidad \'util 
\'unicamente cuando consideramos galaxias por separado. Finalmente, encontramos que 
las regiones clasificadas como DIG en galaxias Irr parecen tener un comportamiento 
m\'as cercano al de las regiones \hii que al DIG en galaxias espirales.}

\abstract{Studies of the Diffuse Ionized Gas (DIG) have progressed without providing 
so far any strict criterion to distinguish DIGs from \hii regions. In this work, we 
compile the emission line measurements of 29 galaxies that are available in the 
scientific literature, thereby setting up the first DIG database (\ddb). Making use 
of this database, we proceed to analyze the global properties of the DIG using the 
\lnii/\ha, \loi/\ha, \loiii/\hb and \lsii/\ha lines ratios, including the \ha emission 
measure. This analysis leads us to conclude that the \nii/\ha ratio provides an 
objective criterion for distinguishing whether an emission region is a DIG or an \hii 
region, while the EM(\ha) is a useful quantity only when the galaxies are considered 
individually. Finally, we find that the emission regions of Irr galaxies classified 
as DIG in the literature appear in fact to be much more similar to \hii regions than 
to the DIGs of spiral galaxies.\vspace{0.12cm}}

\addkeyword{astronomical data bases: miscellaneous}
\addkeyword{galaxies: general} 
\addkeyword{ISM: HII regions}
\addkeyword{ISM: lines and bands}
\addkeyword{line: formation}

\RescaleTitleLengths{1.025}

\begin{document}

\maketitle

\section{Introduction}
\label{sec:Intro}

From  observations of free-free continuum absorption, Hoyle \& Ellis (1963) 
inferred the existence of a new gas component of the interstellar medium (ISM). 
They proposed that this new component was warm ($T_e \sim 10^4$ K), of low 
density ($n_e \sim 10^{-1}\, cm^{-3}$), ionized and that it surrounded the Milky 
Way's disk. The integrated luminosity of this hypothetical medium was comparable 
to that expected from photoionization by O and B stars from the disk.


The existence  of this gas component was later confirmed by the signal dispersion 
from pulsars and by the detection of  faint optical  emission lines from the galactic 
diffuse interstellar medium (Reynolds, Scherb, \& Roesler 1973). Deep images of 
external galaxies such as NGC 891 show a similar warm gas component (Dettmar 1990; 
Rand 1998). This component has since been found in several other galaxies (e.g., 
Bland-Hawthorn, Sokolowski, \& Cecil 1991a,b; Veilleux, Cecil, \& Bland-Hawthorn 
1995; Greenawalt, Walterbos, \& Braun 1997; Martin \& Kennicutt 1997; Wang, Heckman, 
\& Lehnert 1997). Today, it carries various names, such as WIM (Warm Ionized Medium), 
DIM (Diffuse Ionized Medium), DIG (Diffuse Ionized Gas) or eDIG (extraplanar DIG), 
and ``Reynold's Layer'' (in the case of the Galaxy). Hereafter, we will use the 
generic names DIG and eDIG.

Despite the fact that the DIG in our Galaxy accounts for $\sim$ 90\% of the ionized 
medium's mass (Reynolds 1991), the nature of this low density plasma is far from 
being well understood. The first problem encountered in DIG studies is how to 
formally distinguish DIGs from \hii regions, which  similarly consist of a warm and 
ionized gas phase. Several methods have been proposed in the literature. Examples 
are the \ha equivalent width (Bland-Hawthorn et al. 1991a), the \ha emission 
measure\footnote{The EM in units of pc\,${\rm cm^{-6}}$ is defined as  
$\mathrm{EM} \equiv \int n_e n_{H^+} \, ds \approx \int n_e^2 \, ds$.}, EM (e.g., Walterbos 
\& Braun 1994), or the surface brightness (e.g., Ferguson et al. 1996). Nevertheless, 
none of the methods has so far been accepted as standard, mainly because the density 
and size of the emitting regions are usually unknown. To illustrate the above problem, 
an example is provided by the galaxy M31, for which Walterbos \& Braun (1994) adopted 
the EM criterion to distinguish DIGs from \hii regions. They found that close to low 
star formation regions, the DIG always had an EM $<50$\,pc\,${\rm cm^{-6}}$, while 
near higher star formation regions, the latter had an EM near 100\,pc\,$\rm cm^{-6}$. 
Therefore, even though the EM is proportional to the electron density and should thereby 
reflect how ``diffuse'' an emission region is, this criterion  was shown to vary with 
position in this galaxy. 

Another problem in the study of the DIG is the difficulty in observing it, mainly 
because of its low surface brightness. As a result, the observations in the 
literature are sparse and only a few line ratios are typically reported. Furthermore, 
systematical comparative studies of the general behavior of the EMs or of line ratios 
among different objects are still missing. For example, even though a majority of 
authors agree on the existence of an anti-correlation between the \lnii/\ha or 
\lsii/\ha ratios with the EM (these ratios increase when the EM decreases, e.g., 
Bland-Hawthorn et al. 1991a; Rand 1998), very few studies were done with more than 
two galaxies (e.g., Wang et al. 1997; Zurita, Rozas, \& Beckman 2000; T\"ullmann \& 
Dettmar 2000; Miller \& Veilleux 2003). In this work, we will verify for the first time 
whether such an anticorrelation is a general feature of the DIG or not. Another 
interesting aspect is the apparent confusion concerning the \loiii/\hb\ ratio in 
the literature, for which some authors report a constant behavior with EM (e.g., 
Wang et al. 1997, at least for some galaxies), while others report an increase when 
the EM decreases (e.g., Rand 1998; T\"ullmann \& Dettmar 2000).

The spectral characteristics (different from those of \hii regions) combined with 
the large spatial extension of the DIG (typically $|z| \approx 2$~kpc), are not 
well reproduced by models that consider only photoionization from O-B stars (e.g., 
Rand, Wood, \& Benjamin 2009), even though it could provide the entire UV photon 
flux required to mantain the DIG ionized (e.g., Reynolds 1990; Zurita et al. 2000). 
As a result, several authors have suggested that the DIG could be excited by extra 
ionization (or heating) sources like: shocks (e.g., Sivan, Stasi\'nska, \& Lequeux 
1986), mixing layers (e.g., Binette et al. 2009), old hot stars (e.g., Sokolowski 
\& Bland-Hawthorn 1991), and decaying dark matter (e.g., Dettmar \& Schultz 1992). 
In the study of the Seyfert 2 galaxy NGC 1068, Bland-Hawthorn et al. (1991b) proposed 
photoionization by the AGN power law. None of these models is completely successful 
in explaining the line ratios, spatial extensions and luminosities of the DIGs in 
spiral galaxies. For a complete review of the posibles DIG ionizing or heating sources, 
see Bland-Hawthorn, Freeman, \& Quinn (1997). 

With the previous arguments in mind, and to correct for the lack of comparative 
studies in the field of diffuse emission, we have compiled a comprehensive database 
of emission line ratios or fluxes, from DIGs or \hii regions, published in the 
literature. This database is hereafter called the Diffuse Ionized Gas Emission 
Database (\ddb) which will soon be available at the VizieR website of CDS (Centre 
de Donn\'ees Astronomiques de Strasbourg, \url{http://vizier.u-strasbg.fr}).

The structure of the article is the following: after describing \ddb in \S~\ref{sec:DB}, 
we proceed with the analysis and interpretation of the data in \S~\ref{sec:ID}. A 
discussion follows in \S~\ref{sec:discussion} and a brief conclusion is given in 
\S~\ref{sec:conclusions}.


\begin{table*}[!t]
\vspace{0.1cm}
\centering
\setlength{\tabnotewidth}{0.99\textwidth}
\setlength{\tabcolsep}{0.77\tabcolsep}
\tablecols{6}
\caption{Bibliographical references used for the compilation of \ddb}
\label{referencias}
\scriptsize 
\begin{tabular}{cccccc}
\toprule
{\bf Ref\_ID\tabnotemark{a}} &{\bf Author} & {\bf Galaxy} & {\bf Slit\tabnotemark{b}} & {\bf Observations\tabnotemark{c} } & {\bf No. regions} \\ 
\midrule
1& Benvenuti, D'Odorico, \& Peimbert (1976) & M33 & Interarm & \oiii$\lambda$4959, \oiii, \nii & 62\\ 
\noalign{\smallskip}
2& Golla, Dettmar, \& Domgorgen (1996) & NGC 4631 & $\perp$ & \nii, \sii & 39\\ 
 & & & $\parallel$ & \nii, \sii & 80\\ 
\noalign{\smallskip}
3& Greenawalt et al. (1997) & M31 & DIG, \hii & \oii, \oiii, \nii, \siip, EM & 18\\ 
\noalign{\smallskip}
4& Martin \& Kennicutt (1997) & NGC 1569 & DIG & \oiii, \he, \oi, \nii, \siip & 25\\ 
\noalign{\smallskip}
5& Wang et al. (1997) & M51 & Interarm & \oiii, \nii, \siip & 3\\
 & & M101 & Interarm & \oiii, \nii, \siip & 6\\
 & & NGC 2403 & Interarm & \oiii, \nii, \siip & 5\\
 & & NGC 4395 & Interarm & \oiii, \nii, \siip & 3\\ 
\noalign{\smallskip}
6& Rand (1998) & NGC 891 & $\parallel$ &\oi, \nii, \sii& 45\\ 
 & & & $\perp$ & \oiii, \oi, \nii, \sii & 19\\ 
\noalign{\smallskip}
7& Galarza, Walterbos, \& Braun (1999) & M31 & DIG, SBRs & \oii, \oiii, \nii, \siip, EM & 30\\ 
\noalign{\smallskip}
8& Hoopes, Walterbos, \& Rand (1999) & NGC 4631 & $\perp$ & \oiii, \siip, EM& 31\\ 
\noalign{\smallskip}
9& T\"ullmann \& Dettmar (2000) & NGC 1963 & $\perp$ &\oiii, \he, \oi, \nii, \sii, EM& 40\\ 
 & & NGC 3044 & $\perp$ &\oiii, \he, \oi, \nii, \sii, EM& 47\\
 & & NGC 4402 & $\perp$ &\nii, \sii, EM& 24\\
 & & NGC 4634 & $\perp$ &\nii, \sii, EM& 40\\ 
\noalign{\smallskip}
10& Collins \& Rand (2001) & NGC 4302 & $\perp$ & \nii, \siip & 13\\ 
 & & NGC 5775 & $\perp$ & \oi, \oiii, \nii, \siip & 31\\ 
 & & UGC 10288 & $\perp$ & \oi, \oiii, \nii, \siip & 19\\ 
\noalign{\smallskip}
12& Otte, Gallagher, \& Reynolds (2002) & NGC 5775 & $\perp$ & \oii, \oiii, \nii, \siip & 38\\
 & & NGC 4634 & $\perp$ & \oii, \oiii, \nii, \siip & 34\\ 
 & & & $\parallel$ & \oii, \oiii, \nii, \siip & 32\\
 & & NGC 4631 & $\parallel$ & \oii, \oiii, \nii, \siip & 34\\
 & & NGC 3079 & $\parallel$ & \oii, \oiii, \nii, \sii & 53\\
 & & NGC 891 & $\parallel$ & \oii, \nii, \siip & 50\\ 
\noalign{\smallskip}
13& Hoopes \& Walterbos (2003) & M33 & Interarm & \oiii, \he, \nii, \sii,  EM&37\\ 
 & & M51 & Interarm  & \oiii, \nii, \sii, EM & 26\\ 
 & & M81 & Interarm  & \oiii, \nii, \sii     & 2 \\
\noalign{\smallskip}
14& Miller \& Veilleux (2003) & UGC 2092 & $\perp$ &\nii, \sii&4\\
 & & UGC 3326 & $\perp$ & \nii, \sii&6\\
 & & UGC 4278 & $\perp$ & \oiii, \he, \oi, \nii, \sii&13\\
 & & NGC 2820 & $\perp$ & \oiii, \he, \oi, \nii, \sii&13\\
 & & NGC 3628 & $\perp$ & \oiii, \nii, \sii&13\\
 & & NGC 4013 & $\perp$ & \oiii, \he, \oi, \nii, \sii&9\\
 & & NGC 4217 & $\perp$ & \oiii, \oi, \nii, \sii&17\\
 & & NGC 4302 & $\perp$ & \oiii, \he, \oi, \nii, \sii&9\\
 & & NGC 5777 & $\perp$ & \oiii, \nii, \sii&6\\ 
\noalign{\smallskip}
15& Hidalgo-G\'amez (2006) & ESO245-G05 & DIG, \hii & \oii, \oiii, \nii, \sii&7\\ 
 & & Gr8 & DIG, \hii & \oiii, \he, \nii, \sii&19\\ 
\noalign{\smallskip}
16& Voges \& Walterbos (2006) & M33 & DIG &\oiii, \oi, \nii, \sii & 7\\ 
\noalign{\smallskip}
17& Voges (2006) & NGC 891 & $\parallel$ &\oiii, \oi, \nii, \sii, EM& 16\\
 & & & $\perp$ &\oiii, \oi, \nii, \sii, EM& 10\\ 
\noalign{\smallskip}
18& Hidalgo-G\'amez (2007) & DDO53 & DIG, \hii &\oiii, \sii& 26\\
\bottomrule
\noalign{\smallskip}
\tabnotetext{a}{These numbers also serve as bibliographical entries in \ddb.}
\tabnotetext{b}{Slit orientation in each galaxy. The symbols and descriptors 
are described in \S~\ref{sec:descr}.}
\noalign{\smallskip}
\tabnotetext{c}{The line wavelengths are the following: \loii, \loiii, \lhe, 
\loi, \lnii, \lsii  and \lsiimas.} 
\end{tabular}
\vspace{0.2cm}
\end{table*}


\section{The DIG database}
\label{sec:DB}

An extensive research on DIG observations in the literature led us to build 
a comprehensive  database comprising  DIG and \hii region spectroscopic 
observations  of 29 different galaxies (25 spiral galaxies and 4 irregulars). 
This survey contains galaxies with a significant spread in star formation 
rates, H$\alpha$ luminosities, distances, disk inclinations, slit positions 
and slit orientations.


\begin{table*}[!t]
\vspace{0.1cm}
\centering
\setlength{\tabnotewidth}{0.8\textwidth}
\tablecols{6}
\caption{column headers in DIGEDA\tabnotemark{a}}
\label{database}
\footnotesize 
\begin{tabular}{lccccc}
\toprule
{\bf Column} & {\bf 0} & {\bf 1} & {\bf 2} & {\bf 3} & {\bf 4}\\
{\bf Information} & Obs\_ID & Position & \oii$\lambda$3727 & \oiii$\lambda$4363 & H$\beta$\\ 
\noalign{\medskip}
{\bf Column} & {\bf 5} & {\bf 6} & {\bf 7} & {\bf 8} & {\bf 9}\\
{\bf Information} & \oiii$\lambda$4959  & \oiii$\lambda$5007 & \he$\lambda$5876 & \oi$\lambda$6300 & \nii$\lambda$6548\\ 
\noalign{\medskip}
{\bf Column} & {\bf 10} & {\bf 11} & {\bf 12} & {\bf 13} & {\bf 14}\\
{\bf Information} & H$\alpha$ & \nii$\lambda$6583  & \sii$\lambda$6716 & \sii$\lambda$6731 & \sii$\lambda$6716+6731\\ 
\noalign{\medskip}
{\bf Column} & {\bf 15} & {\bf 16} & {\bf 17} & {\bf 18} & {\bf 19}\\
{\bf Information} &  $\rm {T_e\; (10^4\, K)}$ & \sii$\lambda$6716/6731 & \oiii$\lambda$5007/4959  &  H$\alpha$/H$\beta$ &EM\\ 
\noalign{\medskip}
{\bf Column} & {\bf 20} &  {\bf 21} & {\bf 22} & {\bf 23} & {\bf 24}\\
{\bf Information} & Ref\_ID & Morphology & Slit & Region\_ID & Gal\_ID \\ 
\bottomrule
\tabnotetext{a}{See \S~\ref{sec:descr} for further details about the information 
contained in the various entries.}
\end{tabular}
\vspace{0.35cm}
\end{table*}


\subsection{Internal data structure of \ddb}
\label{sec:descr} 

Table~\ref{referencias} lists the 17 bibliographic references used in \ddb, 
which amounts to a total of 1061 line measurement data sets. Table~1 also 
summarizes the  main characteristics of the observations contained in each 
reference. The table~is structured as follows: Column\,1 attributes a unique 
identification number to each  reference, which is later used as a reference 
label in \ddb (Ref\_ID),  Column\,2  contains the bibliographic references, 
Column\,3 lists the names of the galaxies studied  in each reference, Column\,4 
is the slit orientation with respect to the galactic plane in the case of 
edge-on galaxies: parallel ($\parallel$) or perpendicular ($\perp$), while 
for the case of face-on galaxies, a short descriptor of  the region  is used 
instead: Interarm, DIG (Diffuse Ionized Gas), \hii (\hii region), SBR (Super 
Bubble Remnant). In  Column\,5, we detail the set of emission lines reported 
for each galaxy while in Column\,6 we give the total number of areas or 
positions\footnote{Throughout the paper, we also refer to these positions as 
``regions''.} observed within each galaxy. There are two factors that may 
affect the accuracy of the line ratios: dust reddening and the presence of 
underlying absorption lines. We have not performed any reddening correction 
although we did use the reddening-corrected ratios when made available by the 
authors. In any case, all line ratios considered in this Paper involve lines 
that are relatively close in wavelengths and are therefore not very sensitive 
to reddening. As for the presence of underlying absorption lines, this is a 
concern only for \ha and \hb. The equivalent width of either absorption line 
is estimated to be  $\sim -2$\,\AA\ (Antonio Peimbert, private communication). 
Since the emission line equivalent widths are {\it not} available from the 
works cited in Table~\ref{referencias}, we could not correct the ratios for 
this effect nor evaluate precisely its impact.  Further details about the data 
reduction performed by each author can be found in the corresponding bibliographical 
entry listed in Table~\ref{referencias}. Hereafter, we will consider the symbols 
\nii, \oi, \oiii, \sii, and \siip,  as referring to the \lnii, \loi, \loiii, 
\lsii and \lsiimas emission lines, respectively.


\begin{table*}[!t]
\vspace{0.12cm}
\centering
\setlength{\tabnotewidth}{0.8\textwidth}
\tablecols{6}
\caption{List of galaxies contained in \ddb}
\label{galaxies}
\footnotesize
\begin{tabular}{cccccc} 
\toprule
{\bf Gal\_ID\tabnotemark{a}} & {\bf Galaxy} & {\bf R.A.} & {\bf DEC.} & {\bf Morphology Type} & {\bf Inclination}\\
\midrule
 & & (h m s) & (d m s) & & (degrees)\\ 
1 & M 31 &00 42 44.3 &+41 16 09 & Sb, {\sc liner} & $77.5$ \\  
2 & M 33 &01 33 50.9 &+30 39 36 & Scd, \hii &$56$\\ 
3 & M 51 &13 29 55.7 &+47 13 53 &Sbc & $64$\\  
4 & M 81 &09 55 33.2 &+69 03 55 &Sab;{\sc liner}  Sy1.8 & $58$\\  
5 & M 101 &14 03 12.6 &+54 20 57 & SBcd & $17$\\  
20 & NGC 891 &02 22 33.4 &+42 20 57 & Sb,  \hii & $64$\\ 
21 & NGC 1569 &04 30 49.0 &+64 50 53 & Irr & \\ 
22 & NGC 1963 &05 32 16.8 &$-$36 23 55 & Scd & $85$\\  
23 & NGC 2403 &07 36 51.4 &+65 36 09 & SBcd, \hii & $62$\\ 
24 & NGC 2820 &09 21 45.6 &+64 15 29 & SBc & $90$\\  
25 & NGC 3044 &09 53 40.9 &+01 34 47 & SBb & $84$\\ 
26 & NGC 3079 &10 01 57.8 &+55 40 47 & SBc, {\sc liner}, Sy 2 & $60$\\  
27 & NGC 3628 &11 20 17.0 &+13 35 23 & Sb & $87$\\  
28 & NGC 4013 &11 58 31.4 &+43 56 48 & Sb, \hii {\sc liner} & $90$\\   
29 & NGC 4217 &12 15 50.9 &+47 05 30 & Sb & $86$\\ 
30 & NGC 4302 &12 21 42.5 &+14 35 54 & Sc & $90$\\  
31 & NGC 4395 &12 25 48.8 &+33 32 49 & Sm, {\sc liner}, Sy 1.8 & $38$\\  
32 & NGC 4402 &12 26 07.5 &+13 06 46 & Sb & $74$\\  
33 & NGC 4631 &12 42 08.0 &+32 32 29 & SBb & $85$\\   
34 & NGC 4634 &12 42 40.9 &+14 17 45 & Sc & $83$\\  
35 & NGC 5775 &14 53 57.6 &+03 32 40 & Sb & $86$\\  
36 & NGC 5777 &14 51 17.8 &+58 58 41 & Sb & $83$\\  
70 & UGC 2092 &02 36 31.6 &+07 18 34 & Scd & $86$\\ 
71 & UGC 3326 &05 39 37.1 &+77 18 45 & Scd & $90$\\  
72 & UGC 4278 &08 13 58.9 &+45 44 32 & SBd & $90$\\  
73 & UGC 10288 &16 14 24.8 &$-$00 12 27 & Sc & $87$\\ 
90 & ESO245 - G05 &01 45 03.7 &$-$43 35 53 & Irr &\\  
91 & Gr 8 &12 58 40.4 &+14 13 03 & Irr & \\  
92 & DDO 53 &08 34 07.2  &+66 10 54 & Irr & \\  
\bottomrule
\noalign{\smallskip}
\tabnotetext{a}{Unique galaxy identification number within \ddb.}
\end{tabular}
\vspace{0.64cm}
\end{table*}


The 1061 observations obtained from these references were extracted by 
digitalization of  published figures or tables. The data were subsequently 
normalized and incorporated in \ddb. This resulted in a table of 25 
comma-separated columns (csv format) containing 1061 data lines (or records). 
Missing entries are represented by ($-1$) in the corresponding data field. 
The table begins with a header line, which attributes a descriptor to each 
column. The first column attributes a unique identifying number (Obs\_ID) 
to the observation it refers to. Apart from line ratios or fluxes, additional 
columns (20 to 25) are provided that describe the particularities of the 
observation (observed position, slit orientation, ...). The entries contained 
in these columns consist of numbers (or pointers), which refer to specifically 
coded information (see description  below). This ensures that the information 
it contains can be easily accessed by any numerical program. A list of all the 
column headers and the type of information they contain is provided in Table~\ref{database}.

As we mentioned above, the entries in some of the columns have a numerical 
format that deserves further clarification: ``Obs\_ID'' is the positive, 
consecutive and unique number for row identification. ``Position'' is the 
value in kpc of the position within each galaxy. In case the observations 
was carried out perpendicular to the galaxy plane, this is the distance 
from the plane, for parallel slits this is the position from the galaxy 
center and for face-on galaxies the position is determined with respect to 
some reference point that the authors had specified. The ratio H$\alpha$/H$\beta$ 
is that reported by the authors. However, in cases where no value was reported, 
a theoretical ratio of 2.86 was assumed when its value was required to normalize 
the forbidden lines. The ``EM'' is the \ha emission measure, specified in 
units of pc\,${\rm cm}^{-6}$. ``Ref\_ID'' is the integer descriptor that 
identifies the bibliographical reference (see Table~\ref{referencias}). 
``Morphology'' refers to the galaxy's morphology and its orientation, as 
follows: for spiral galaxies  11 is reserved for edge-on objects and 12 for 
face-on ones; 2 is used for irregulars. ``Slit'' is the slit orientation with 
respect to the galactic plane: 1 for parallel slits, 2 for perpendicular slits 
and 3 for observations of face-on objects. ``Region\_ID'' refers to  the three 
possible emission region types: 1 and 11 are used for \hii regions, 2 and 21 
for transition zones, 3 and 31 for DIG regions, and finally,  4 is reserved for 
data that do not fit any classification. Single digit entries denote the authors 
original classification (when it was reported) while double digit entries denote 
our own classification taking into account the value of the height above the 
galactic plane $|z|$ as described below in \S~\ref{sec:z}. Finally, ``Gal\_ID'' 
is  an integer number that relates uniquely the galaxy's name to its source catalog, 
as follows: Messier: 1--19, New General Catalog: 20--69, Uppsala General Catalog: 
70--89 and others: 90--99. The list of all the galaxies in \ddb, with their basic 
characteristics, is given in Table~\ref{galaxies}.

To sum up, \ddb contains 1061 spectroscopic observations. It is worth noting, 
however, that only 114 observations contain simultaneous measurements of \nii, 
\oiii, \sii\ and the EM (these turn out to be edge-on spiral galaxies). But if 
we consider instead a single pair of lines, for example \nii and \oiii, the 
number of available data entries in this case increases to 462 observations. 

\ddb will be made available on-line, at the VizieR website (\url{http://vizier.u-strasbg.fr}).

\subsection{An additional grouping criterion: $|z|$}
\label{sec:z}

Due to the fact that there is no standard criterion to distinguish DIGs from 
\hii regions, most of the authors in the field report their spectroscopic 
studies without any classification. For instance, $\simeq 76 \%$ of the 1061 
observations in the \ddb are  neither referred to as \hii region, DIG or 
transition zone. About $\sim 97 \%$ of these data correspond to edge-on galaxy 
observations,  which are of particular interest among diffuse ionized gas studies. 
For this galaxy type, the halo gas is easier to separate from the disk (where 
most \hii regions are located). In that case, the DIG emission can be reasonably 
assumed to be free from contamination by star formation emission. Considering 
that in the Milky Way, the thin disk is defined as the region where most of 
the galaxy's star formation is taking place, we propose in this Paper a `zoning' 
classification criterion based on the height (above or below) the galactic plane 
$|z|$. The proposed zoning classification leads to the following three situations:

\begin{itemize}
\item $|z| < 500$\,pc. Owing to the fact that in the Milky Way, one considers 
that \hii regions are confined to a disk of 350\,pc height, we have assumed 
this to be a general behavior in  spiral galaxies. We will therefore consider 
that line emission from regions within 500\,pc from the galactic plane corresponds 
overwhelmingly to emission from star forming regions. Consequently, we will 
classify these regions as {\it \hii regions} (``Region\_ID''=11).
\item $500 \leq |z| < 1000$\,pc. The majority of spiral galaxies from the database 
do not have an inclination of exactly $90\arcdeg$ and it is therefore probable that, 
due to projection effects, a line of sight below or above the galactic plane of 
between 500 and 1000\,pc will cross a large volume, which likely will contain both 
DIG and \hii region emission. Since we do not know  precisely the contribution from 
each type, we will classify these regions as being from {\it transition zones} 
(``Region\_ID''=21).
\item $|z| \geq 1$\,kpc. Even though some authors have reported the existence of 
\hii regions as far away as 1.5\,kpc from the plane, this remains an uncommon 
occurrence. In this work, we will assume that the emission from any lines of sight 
above $|z| = 1\,\mathrm{kpc}$ from the galactic plane is free from any contamination 
from star formation emission. This leads us to classify these regions as {\it eDIG} 
(``Region\_ID''=31).
\end{itemize}

Applying this `zoning' classification to the database resulted in 309 \hii regions, 
218 transition zones, and 509 eDIGs. Only 2\% of the observations could not be 
classified. They turn out to come from face-on galaxies where our criterion based on 
$|z|$ is not applicable.


\begin{figure*}[!t]
\makebox[0pt][l]{\textbf{(a)}}%
\hspace*{\columnwidth}\hspace*{0.75\columnsep}%
\textbf{(b)}\\[-0.7\baselineskip]
\parbox[t]{\textwidth}{%
\vspace{0pt}
\includegraphics[width=0.98\columnwidth]{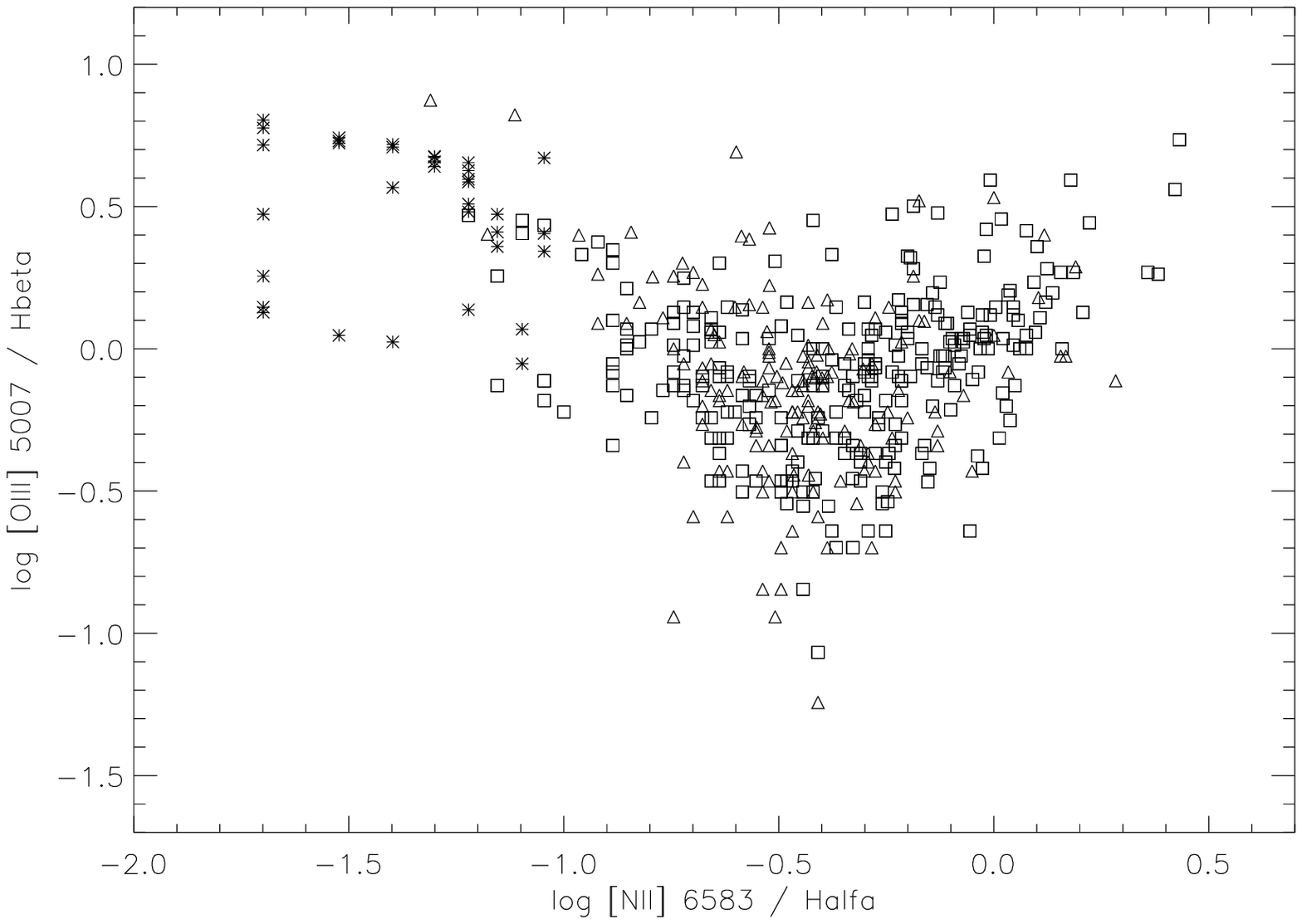}%
\hfill%
\includegraphics[width=0.98\columnwidth]{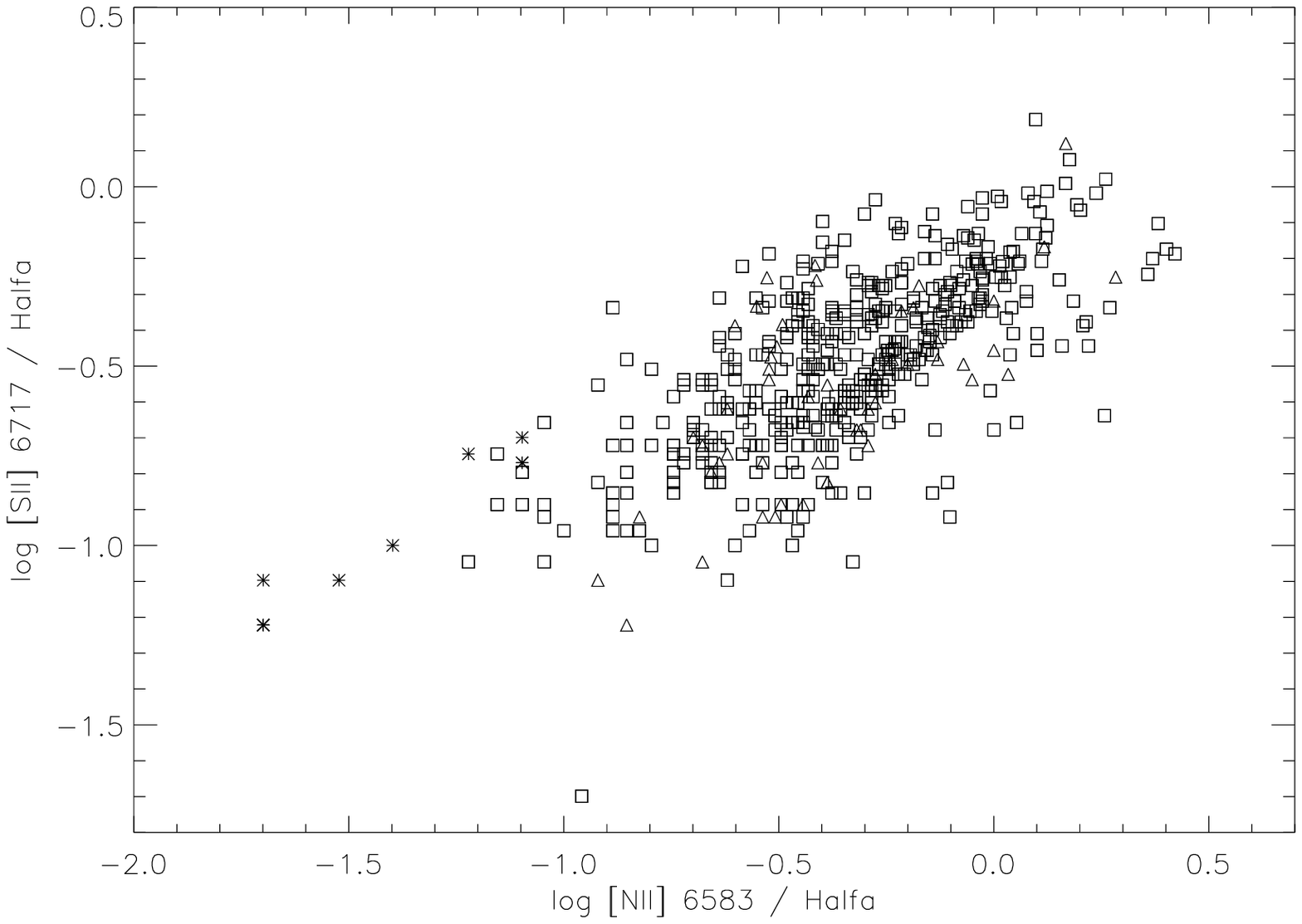}}
\makebox[0pt][l]{\textbf{(c)}}%
\hspace*{\columnwidth}\hspace*{0.75\columnsep}%
\textbf{(d)}\\[-0.7\baselineskip]
\parbox[t]{\textwidth}{%
\vspace{0pt}
\includegraphics[width=0.98\columnwidth]{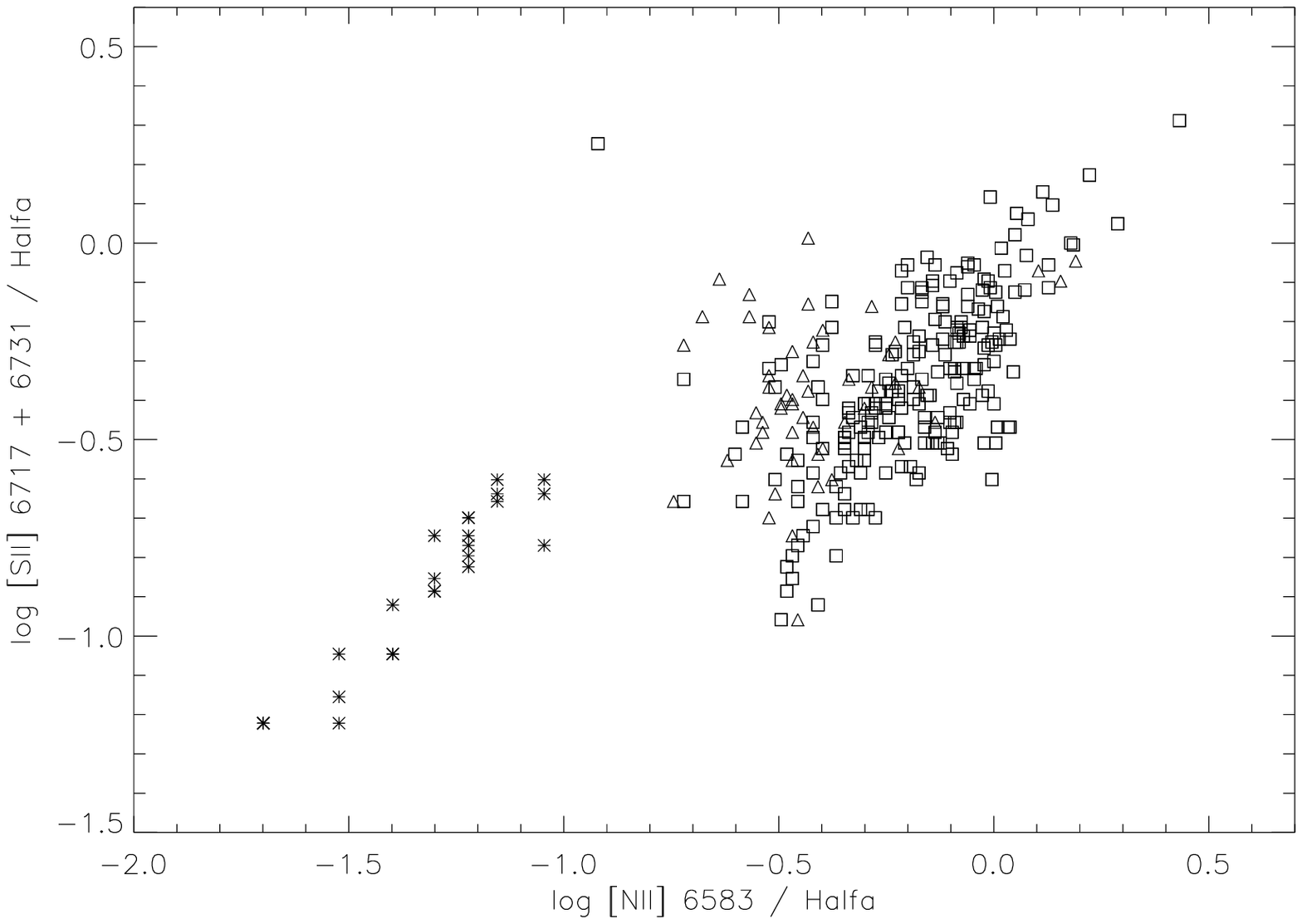}%
\hfill%
\includegraphics[width=0.98\columnwidth]{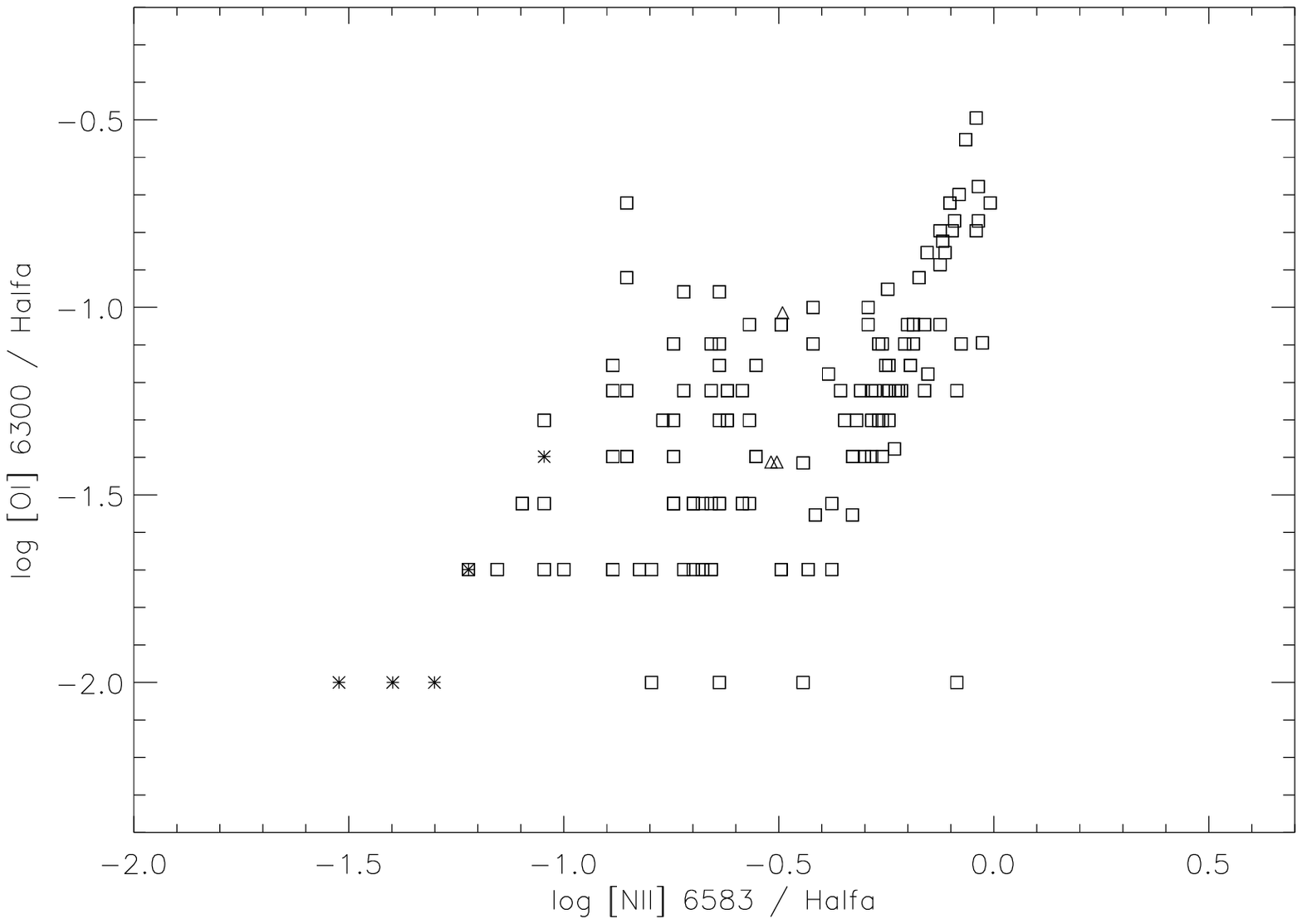}}
\caption{Diagnostic diagrams for all observations in the database. The data points 
identify the galaxy types as follows: irregular galaxies (asterisks), face-on spiral 
galaxies (triangles), and edge-on spiral galaxies (squares).} 
\label{fig:dd_galaxies}
\vspace{0.2cm}
\end{figure*}


\section{Analysis and interpretation of the \ddb data}
\label{sec:ID}

We have carried out a statistical study of the database with the aim of 
identifying the main differences between DIGs and \hii regions. To assist 
us, we used the line ratio diagnostic  diagrams prescribed by Baldwin, 
Phillips, \& Terlevich (1981)  and  by Veilleux \& Osterbrock (1987). Such 
diagrams were created with the intention of classifying line emitting regions 
according to their physical conditions, yet using only the most commonly 
measured lines, that is: \ha, \hb, \nii, \oi, \oiii, and \sii. Baldwin et 
al. (1981)\footnote{Hereafter, we will refer to the \nii/\ha vs. \oiii/\hb 
diagram as `BPT diagram'.} and Veilleux \& Osterbrock (1987) found that 
these diagrams provide a quantitative classification scheme of extragalactic 
sources, in which the main excitation mechanism operating on the gas is 
identified, namely: photoionization by O and B stars, photoionization by a 
power-law continuum source, shock-wave heating and photoionization by old 
and very hot stars. Apart from being used for the above purpose, these diagrams 
also turn out to be useful in other situations or for other kinds of emission 
objects (e.g., distinguishing planetary nebulae from \hii regions, see Kniazev, 
Pustilnik, \& Zucker 2008). Since the emission lines quoted above are among 
the most commonly observed lines in DIG regions as well, we adopted the line-ratio 
diagnostic diagrams in our statistical study.

\subsection{Results from line ratio diagnostic diagrams}
\label{BPT}

In Figures~\ref{fig:dd_galaxies} and \ref{fig:dd_regions}, we show, for the 
whole database, the line ratios \oiii/\hb, \oi/\ha, \sii/\ha, and \siip/\ha\ 
as a function of the ratio \nii/\ha. Figure~\ref{fig:dd_galaxies} distinguishes 
the different galaxy types as follows: Irregular galaxies (asterisks), spiral 
face-on galaxies (triangles) and edge-on spirals (squares). Figure~\ref{fig:dd_regions}, 
which is discussed below, distinguishes among the three emission region types 
defined in \S~\ref{sec:descr}.

Our Panel\,(a) in Figure~\ref{fig:dd_galaxies} shows that the data describes a 
``seagull'' shape quite similar to that found by Stasi\'nska et al. (2006) from 
a study of 20,000 ``normal'' star-forming and AGN galaxies extracted from the 
{\sc sloan} database. While the line ratios from both spiral types (edge-on and 
face-on) extend over the full seagull area in Panel\,(a) of Figure~\ref{fig:dd_galaxies}, 
those from irregular galaxies tend to occupy only the extreme left wing (lower 
\nii/\ha and greater \oiii/\hb values). In the other Panels~(b), (c) and (d) of 
Figure~\ref{fig:dd_galaxies}, the same remarkable behavior is found. Notwithstanding 
the reasons for such behavior (see \S~\ref{sec:discussion}), it is statistically 
apparent that the data from irregular and spiral galaxies should not be analyzed 
without differentiating the galactic type. Since \ddb contains only nine observations 
of irregular galaxies, we chose to consider only spiral galaxies in the analysis 
that follows.

Figure~\ref{fig:dd_regions} is similar to the previous figure, except that it 
distinguishes among the three types of emission line regions: \hii regions (blue 
diamonds), transition zones (green crosses) and DIGs (red pluses). In Panel\,(a), 
DIGs and \hii regions conform together a ``seagull'' shape. However, it becomes 
evident in this BPT diagram that the DIGs occupy a very different location with 
respect to the \hii regions. While \hii regions are principally located in the 
left seagull wing, DIGs cover the right seagull wing, i.e. the two types can be 
distinguished from their \nii/\ha ratio alone.

Although less apparent, we can see in Panels~(b) and (c) of Figure~\ref{fig:dd_regions} 
that DIGs tend furthermore to have \sii/\ha ratios greater than those from 
\hii regions. This behavior has been previously mentioned on several occasions 
(Bland-Hawthorn et al. 1991a; Zurita et al. 2000; Rand 1998; Haffner, Reynolds, 
\& Tufte 1999; T\"ullmann \& Dettmar 2000) for different galaxies using different 
classification criteria. The importance of this figure in DIG studies is that 
it makes clear that even when the data have been classified using a very different 
criterion, the intensities of \nii and \sii are significantly greater in general 
in DIGs than in \hii regions. Perhaps the spread in position of each type and 
the extent to which they overlap (which differ in each panel) could also be 
the manifestation of the absence of an universally accepted classification 
scheme to distinguish the emission region type.

Conspicuously, Panel\,(d) of Figure~\ref{fig:dd_regions} does not show any 
clear trend in the behaviour of the \oi/\ha line ratio. The data dispersion 
and the overlap in this panel are both larger than found with \nii or \sii, 
possibly because the \oi line is the weakest of all lines considered here. 
In this case, we can expect the \oi measurements to be more difficult and 
uncertain, which is consistent with the fact that we only have 148 \oi 
measurements (all types together) available among the 1061 observations.

\subsection{The \nii/\ha line ratio criterion}
\label{NIIcriteria}

\begin{figure*}[!t]
\vspace{0.1cm}
\makebox[0pt][l]{\textbf{(a)}}%
\hspace*{\columnwidth}\hspace*{0.75\columnsep}%
\textbf{(b)}\\[-0.7\baselineskip]
\parbox[t]{\textwidth}{%
\vspace{0pt}
\includegraphics[width=0.98\columnwidth]{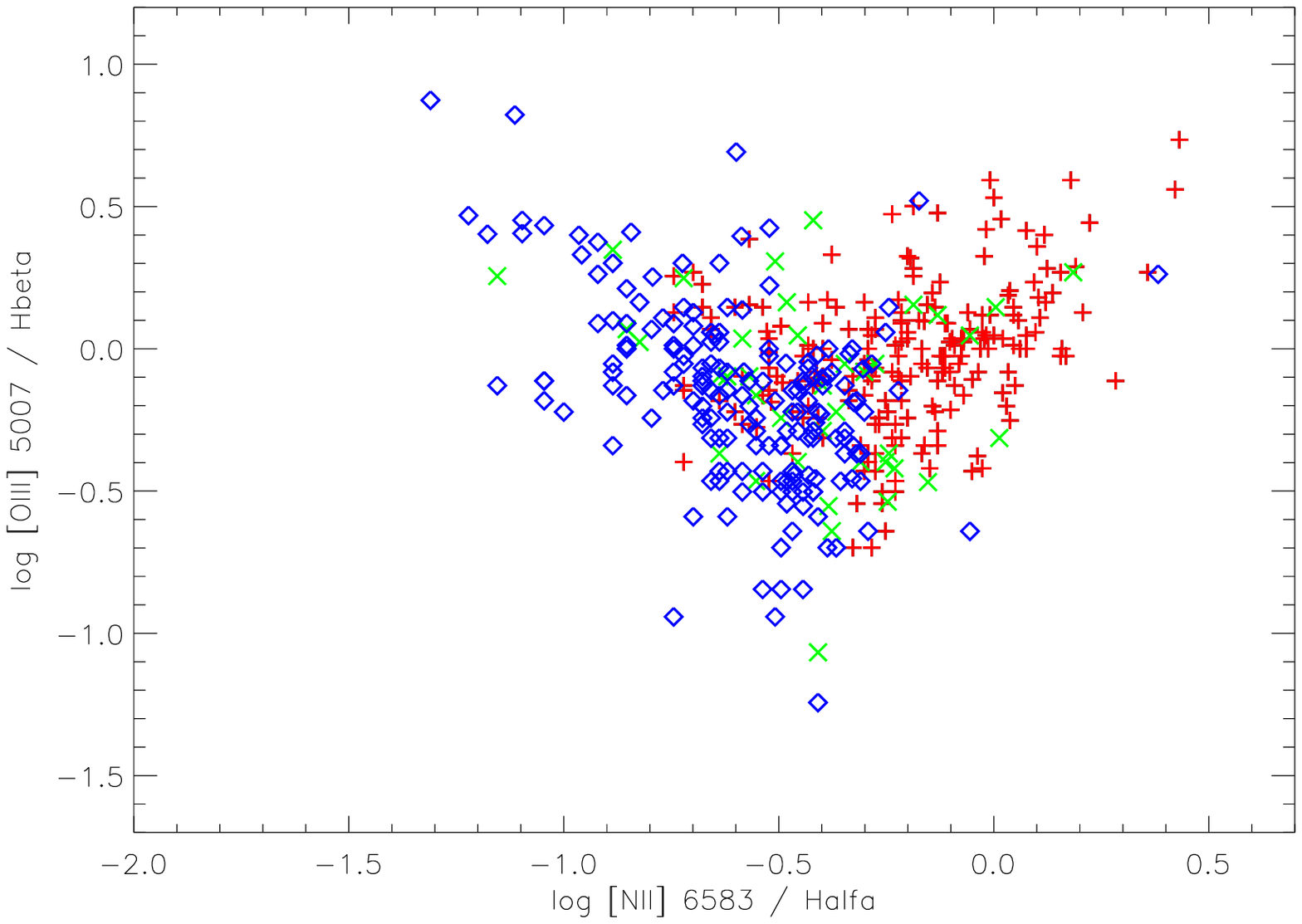}%
\hfill%
\includegraphics[width=0.98\columnwidth]{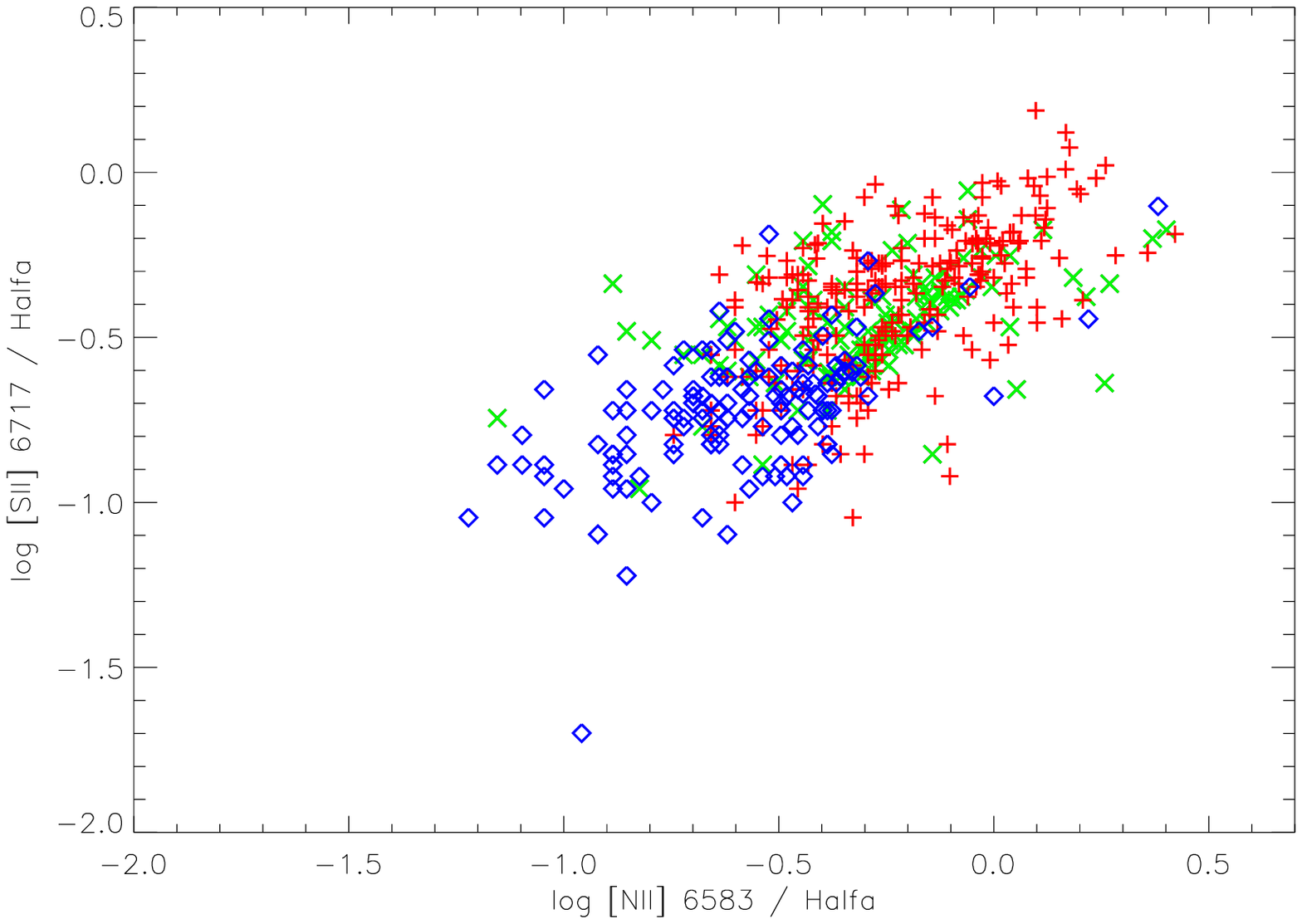}}
\makebox[0pt][l]{\textbf{(c)}}%
\hspace*{\columnwidth}\hspace*{0.75\columnsep}%
\textbf{(d)}\\[-0.7\baselineskip]
\parbox[t]{\textwidth}{%
\vspace{0pt}
\includegraphics[width=0.98\columnwidth]{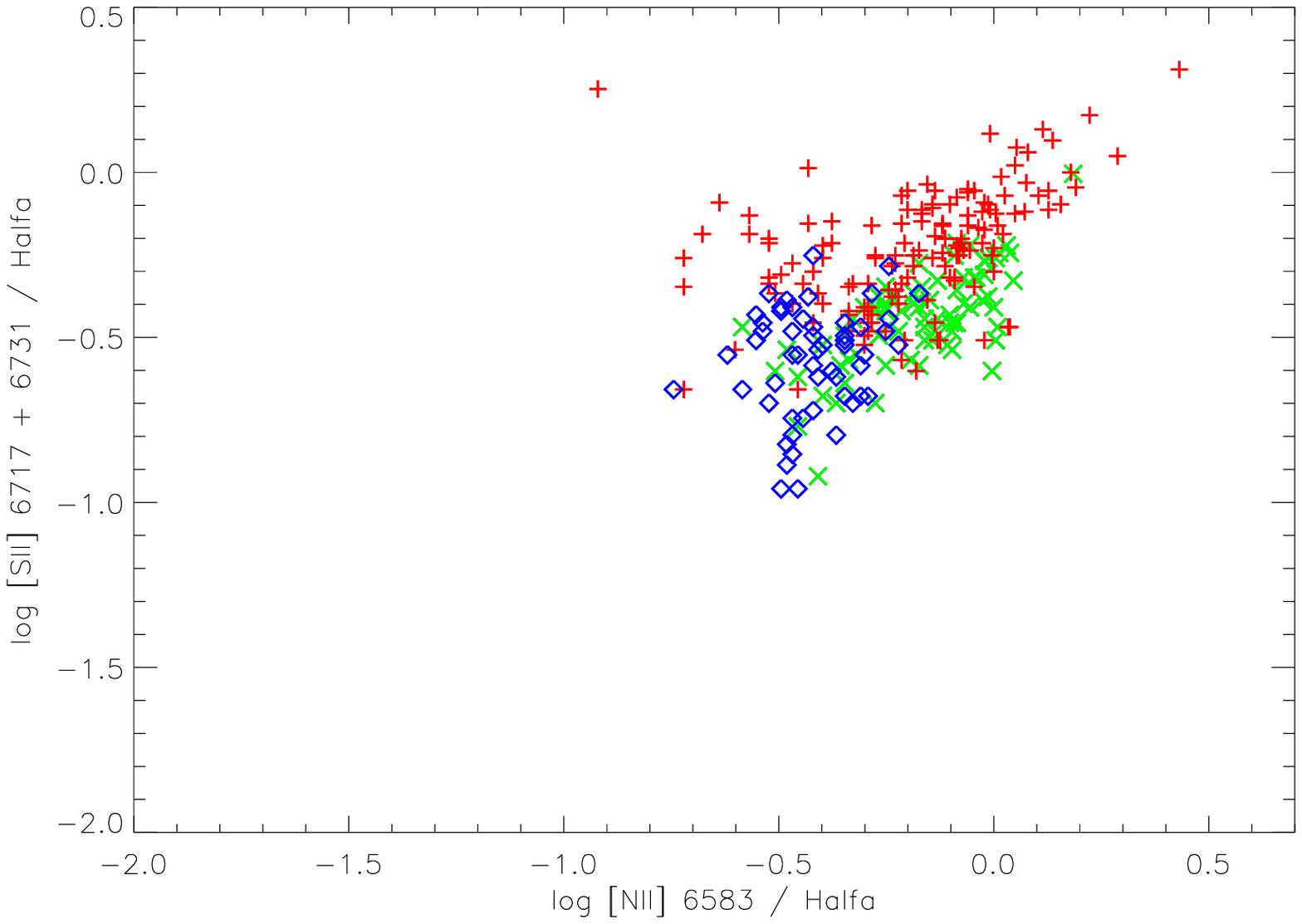}%
\hfill%
\includegraphics[width=0.98\columnwidth]{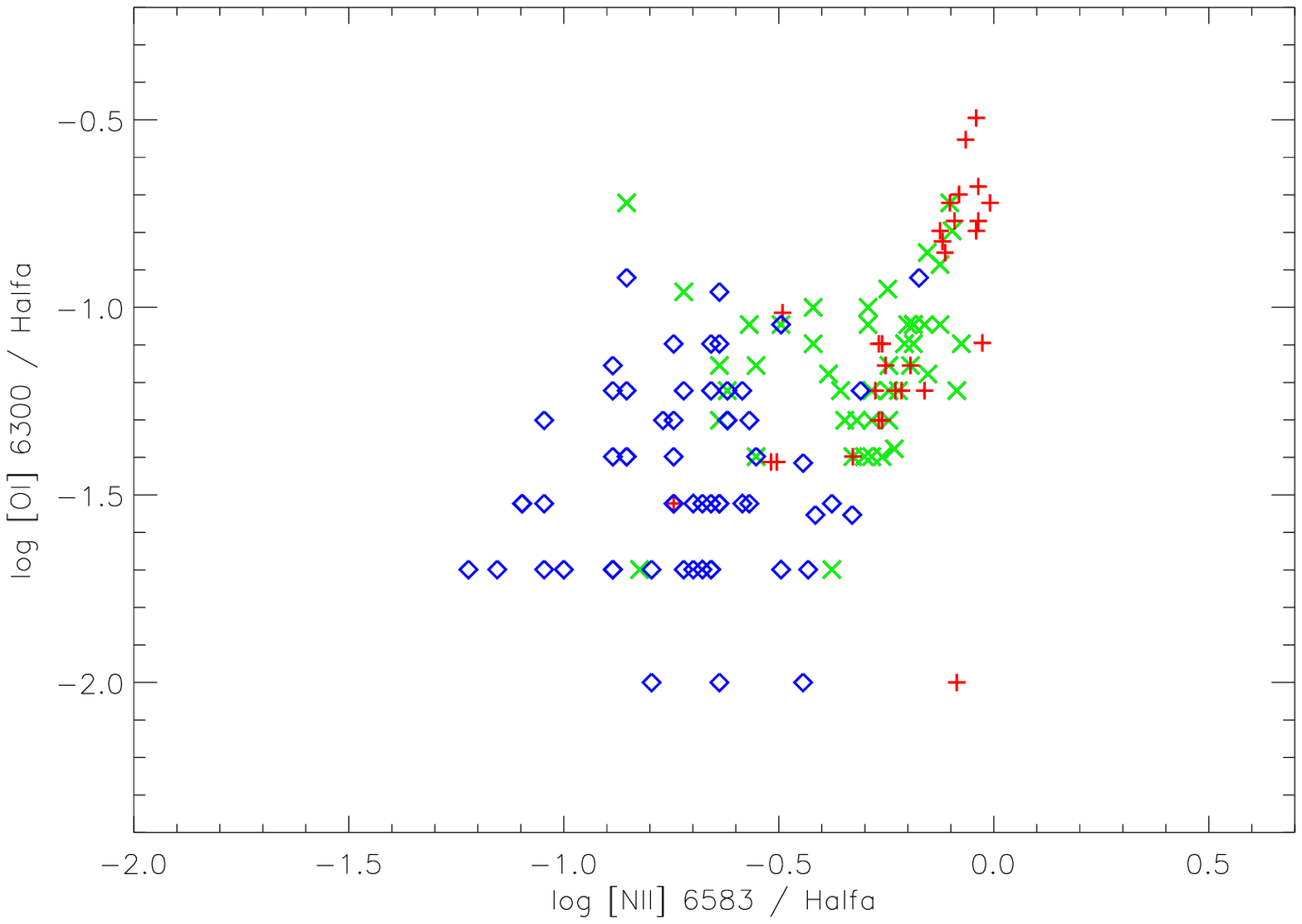}}
\caption{Diagnostic diagrams for all observations in the database. The data 
points identify the emission region types as follows: \hii regions (blue 
diamonds), transition zones (green crosses) and DIGs (red pluses).} 
\label{fig:dd_regions}
\vspace{0.2cm}
\end{figure*}

Even though our diagnostic diagrams in Figure~\ref{fig:dd_regions} strongly 
suggest the existence of a clear separation between \hii regions and DIGs in 
the behavior of \nii/\ha and \sii/\ha, a significant overlap is nevertheless 
present. It thus becomes essential to make a quantitative comparison of the 
distributions of the various line ratios. This exercise would allow us to 
identify which ratios show the most significant differences between DIGs and 
\hii regions in the distribution of their values. For this purpose, we can 
use box diagrams, which summarize in a very schematic fashion the intrinsic 
distribution of any of the data entries. In such diagrams, each box contains 
50\% of the data and each has been defined using quartiles Q1 and Q3, which 
are the lowest and highest 25\% cuts-offs in the data distribution. In such 
boxes, a vertical line denotes the position of the median of the distribution, 
while ``whiskers'' are used to represent distribution boundaries that include 80\% 
of the data, which have been defined using the deciles D1 and D9, corresponding 
to the lowest and highest 10\% cut-offs.

\begin{figure}[!t]
\centering
\includegraphics[width=8cm,height=8cm]{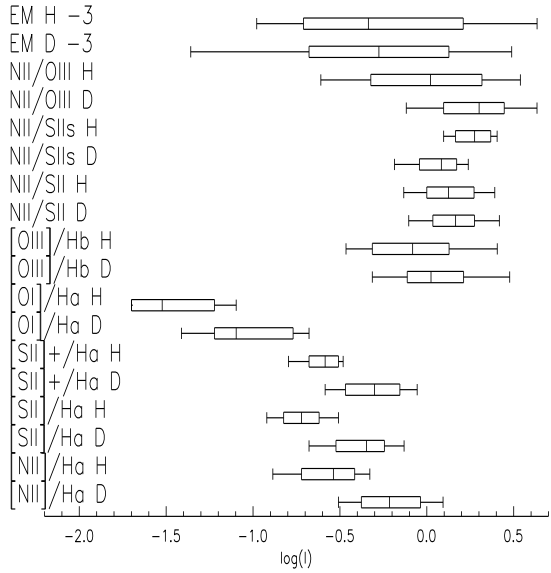}
\caption{Box diagrams of the data distributions of the EM and of 
the most prominent line ratios from DIGs and \hii regions (D and 
H labels, respectively). All the values are logarithmic, with that 
of the EM reduced by 3 dex.}
\label{fig:bd}
\vspace{0.25cm}
\end{figure}

Figure~\ref{fig:bd} shows such box diagrams for the EM values and for each 
of the most prominent line ratios (from spiral galaxies) contained in \ddb. 
Each datum's statistic distinguishes whether it considers only DIGs or \hii 
regions, using the label (D) or (H), respectively. It is apparent that a few 
conspicuous line ratios show a distribution that is significantly {\it distinct} 
between DIGs and \hii regions. These ratios correspond to those boxes that 
do not overlap (or overlap slightly) in log (I) position when comparing the 
(D) and (H) boxes of the same quantity. It is also obvious that, conversely, 
many line ratios show an important overlap in the position of the (D) and (H) boxes.

If we consider the line ratios of \oi, \sii, \siip and \nii  with respect to 
\ha, the following sample sizes are available: \oi (157 entries), \sii (521 
entries), \siip (322 entries)  and \nii (888 entries). It is noteworthy that 
for each of these 4 line ratios, the distributions that the boxes in 
Figure~\ref{fig:bd} represent do {\it not} overlap between the (D) and (H) 
entries. That is, these 4 line ratios show clear differences in their quantiles 
distribution between DIGs and \hii regions. This quantitative analysis gives 
us further confidence in proposing that a few line ratios are by themselves 
sufficient to empirically distinguish  a DIG region from an \hii region, 
notwithstanding the lack of any prior knowledge about the electron density or 
column size (both quantities being unknown in DIGs).

Even considering that any of the above four lines ratios could be a useful 
criterion for distinguishing DIGs, the \nii/\ha statistic corresponds to the 
largest data sample. Observationally, it is also perhaps  the ratio that is 
easiest to measure (since \nii is a strong line and the \nii/\ha ratio does 
not require any reddening correction).

In Table~\ref{quantiles}, the numerical values of the different quantiles for 
\nii/\ha are sumarized. It can be concluded that emission regions with \nii/\ha 
$< -0.5$ should be classified as \hii regions, while those with \nii/\ha $> -0.3$ 
should be classified as DIGs. Measurements of intermediate values result in an 
uncertain classification. In any case,  we can infer that emission regions with 
a ratio \nii/\ha $< -0.4$ have a high probability of being an \hii region while 
regions with \nii/\ha $> -0.4$ have a higher probability of being a DIG.

Additionally, the boxes in Figure~\ref{fig:bd} representing the ratios 
\oiii/\hb, \nii/\sii, \nii/\siip, and \nii/\oiii do {\it not} show a 
significant difference in distribution between DIGs and \hii regions. 
These ratios are therefore useless for distinguishing one type of region 
from another.

Some authors have proposed to use the EM as a classification criterion 
(e.g., Walterbos \& Braun 1994). We note, however, from Figure~\ref{fig:bd} 
that the box diagrams of EM do not show any significant differences between 
the DIG and \hii region distributions. Other authors (e.g., Bland-Hawthorn 
et al. 1991a,b; Rand 1998) have suggested that the \nii/\ha ratio shows a 
strong anti-correlation with the EM, i.e. the \nii/\ha would increase as 
the EM decreases. To which extent therefore can the EM be a distinguishing 
universal criterion for DIGs? Even though the EM distribution box in 
Figure~\ref{fig:bd} suggests that we cannot (because of a large overlap of 
the (D) and (H) boxes), we may nevertheless consider the possibility that 
within a single galaxy, this criterion might be applicable.  After all, it 
is the only quantity in the box diagrams of Figure~\ref{fig:bd} that does 
not correspond to a ratio and is therefore not normalized. The wide overlap 
might for instance be the result of a wide spread in EM values from galaxy 
to galaxy.

\begin{figure}[!t]
\makebox[0pt][l]{\textbf{(a)}}%
\parbox[t]{\columnwidth}{%
\vspace{0.1cm}
\includegraphics[width=0.98\columnwidth]{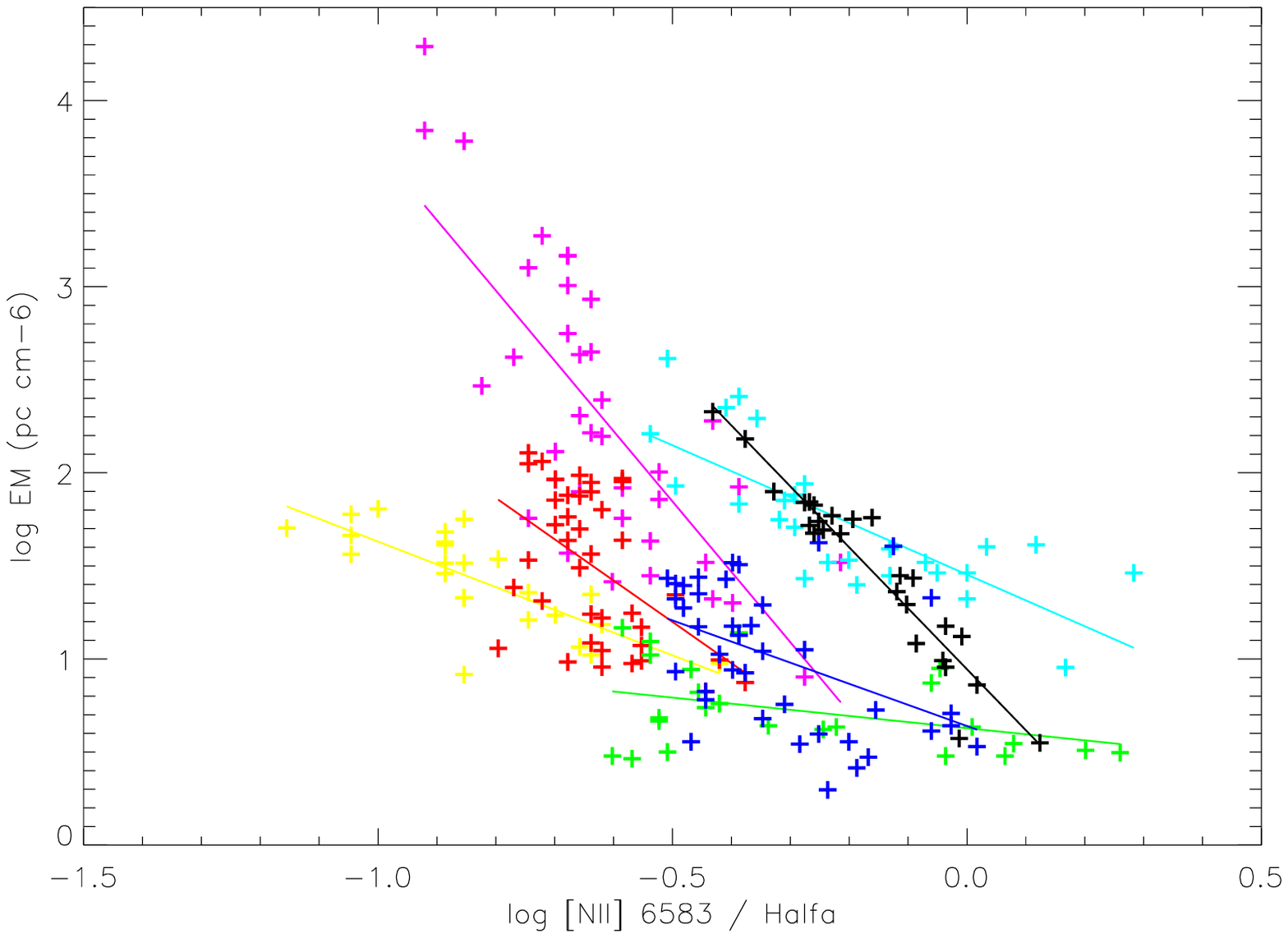}
\vspace{0.2cm}}
\makebox[0pt][l]{\textbf{(b)}}%
\parbox[t]{\columnwidth}{%
\vspace{0.1cm}
\includegraphics[width=0.98\columnwidth]{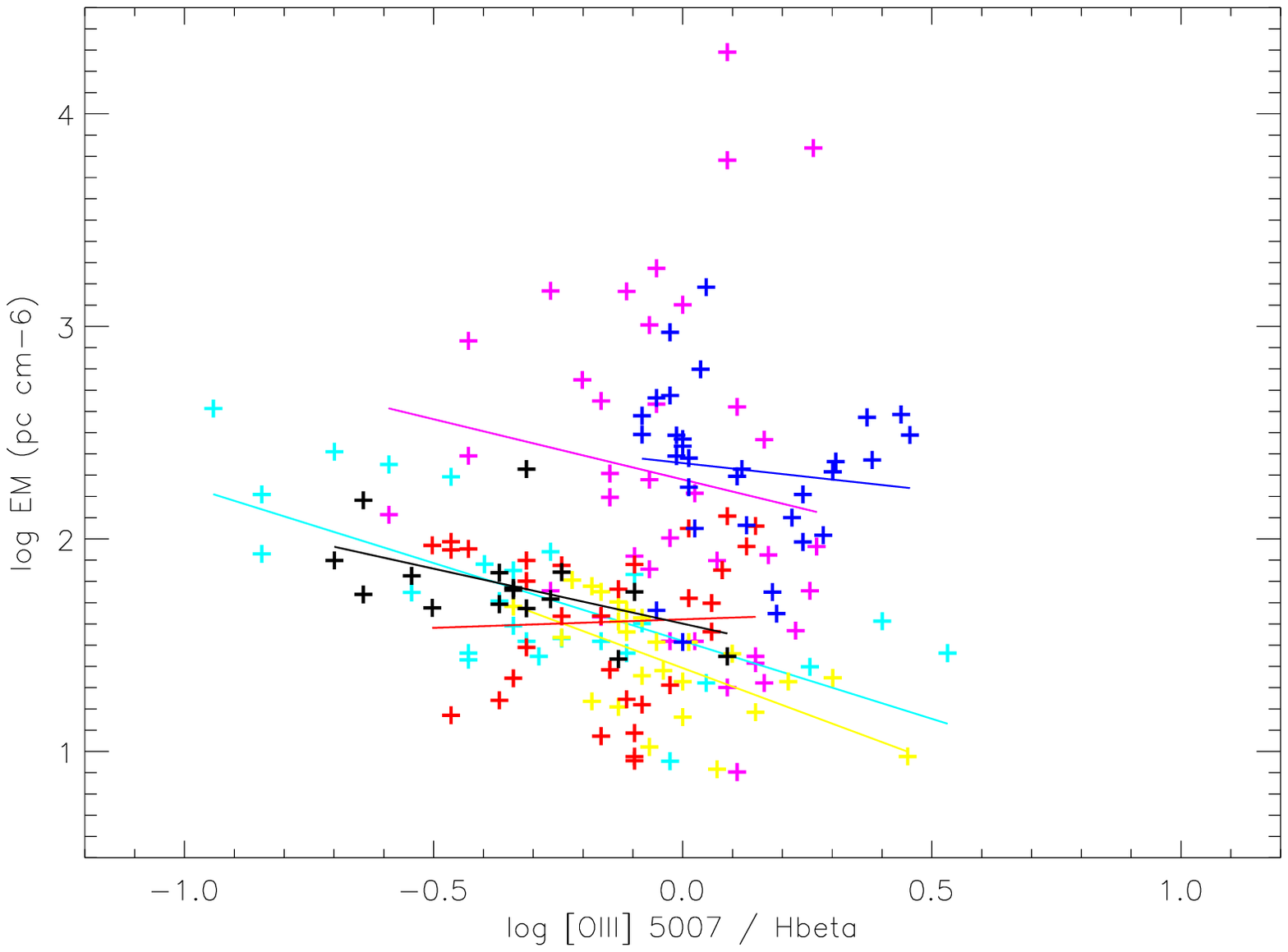}}
\caption{The EM against (a) \nii/H$\alpha$  and (b) \oiii/$\beta$. 
A different color coding is used for each galaxy: M33 (magenta), 
M51 (cyan), NGC 1963 (yellow), NGC 3044 (red), NGC 4402 (green), 
NGC 891 (black) and NGC 4634 (blue, in panel a), NGC 4631 (blue, in panel b). 
The best log-log linear fits of the points of each galaxy are overplotted 
using the same color coding.} 
\label{fig:ajustes}
\vspace{0.25cm}
\end{figure}

In order to explore further this  question, we plotted EM values against 
several emission line ratios. As expected, we found a very feeble general 
relationship and only in the case of \nii/\ha. However, when we considered 
individual galaxies separately, a different picture emerged. Our results are 
summarized in Figure~\ref{fig:ajustes} where Panel~(a) has \nii/\ha in the 
abscissa and Panel~(b) \oiii/\hb, while the EM is displayed in the ordinate. 
A different color coding is used for each galaxy: M33 (magenta), M51 (cyan), 
NGC 1963 (yellow), NGC 3044 (red), NGC 4402 (green), NGC 891 (black), NGC 4634 
(blue, Panel~(a) only), NGC 4631 (blue, Panel~(b) only). The best log-log 
fits of the points of each galaxy are overplotted. In Panel~(a), we see that 
there exists indeed a relationship between EM and \nii/\ha for each galaxy 
{\it individually}. In the case of \oiii/\hb (Panel b), a similar but weaker 
relationship appears to be present as well. 

Even though a general EM-based criterion cannot be defined (Figure~\ref{fig:bd}), 
it is apparent from Figure~\ref{fig:ajustes} that when the \nii/\ha ratio 
(Table~\ref{quantiles}) is used to determine the emission region type, a 
critical value ${\rm EM_c}$ can then be found that appears to be different 
for each galaxy individually.  We obtain the interesting result that emission 
regions with EM $\rm > EM_c$ can be considered to be \hii regions while 
emission regions {\it in the same galaxy} with EM $\rm < EM_c$ turn out to 
be DIGs. This particular ${\rm EM_c}$ criterion, defined for each galaxy, 
is based on the more general \nii/\ha criterion. ${\rm EM_c}$ allows us now 
to distinguish DIGs from \hii regions with certainty, confirming that the 
basic physical conditions (as represented by line ratios) in these two 
region types are indeed different.

\section{Discussion}
\label{sec:discussion}

Following our statistical analysis of the database, there are two results 
that deserve further discussion: on the one hand, the different physical 
conditions taking place in DIGs and \hii regions (in the context of spiral 
galaxies) and, on the other hand, the extreme position that the DIGs from 
irregular galaxies occupy in the BPT-diagram with respect to the DIGs from 
spirals.


\subsection{DIG and \hii regions: what is the underlying difference?}
\label{sec:DIGlocation}

BPT proposed a way to classify emission line regions, which is based on their 
location in the \oiii/\hb vs. \nii/\ha diagnostic diagram. These authors were 
successful in separating \hii regions (low \nii/\ha) from {\sc liners} and 
Seyfert galaxies (high \nii/\ha). Using data from \ddb, we found that most of 
the DIGs fall in the regions occupied by {\sc liners} and Seyferts (see 
Figure~\ref{fig:dd_regions}). However, we cannot foresee how in non-active 
galaxies the same kind of photoionization source that powers AGN could replace 
O--B stars of \hii regions and become responsible for the excitation of the DIG.

More recently, using the SDSS survey Stasi\'nska et al. (2008) proposed an 
original interpetation for objects occupying the right part of the BPT diagram 
(which they dubbed the seagull's right wing). They pointed out  that the 
simultaneous increase of \oiii/\hb and \nii/\ha can also be the result of 
photoionization by hot old stars, for example the central stars of planetary 
nebulae (PNe). According to this interpretation, the ionizing source of the 
DIG would be similar to that of \hii regions (i.e. a stellar continuum), 
except for the effective temperature of the ionizing stars and, consequently, 
the higher electron temperature of the gas.

Following the conclusions reached by Stasi\'nska et al. (2008), we propose 
that the DIG could be ionized by old stars, at least in spiral galaxies. In 
fact, the suggestion that a hot stellar population might contribute to the 
ionizing source was first proposed by Lyon (1975) and later explored by 
Sokolowski \& Bland-Hawthorn (1991) who developed photoionization models that 
combine OB stars with old hot stars. Sokolowski \& Bland-Hawthorn (1991) 
concluded that such composite models can reproduce the DIGs observations in 
NGC 891. However, as early as in 1991, the available observational data were 
scant and no \oiii/\hb ratios from DIGs had yet been published.

One can find other explanations for the  position of DIGs in the BPT diagram 
besides that of a hotter photoionization source. For instance, the photoionization 
models developed by Sokolowski (1993) suggest that the ionizing flux from O 
and B stars could simulate the effect of a hotter source when intervening dust 
is present (between the stars and the DIG) and the selective extinction by 
such dust is taken into account. A similar UV hardening could also result 
from the radiation that leaks out from disk \hii regions (Domgorgen \& Mathis 
1994; Zurita et al. 2000). Both effects of selective absortion and of photoionization 
by old and hot stars are the subject of a paper in preparation.

\subsection{DIG in irregular galaxies}
\label{sec:Irrlocation}

The DIGs observed in irregulars occupy the same part of the BPT-diagram as 
do the \hii regions of spiral galaxies (low \nii/\ha in Figure~\ref{fig:dd_galaxies}). 
Furthermore, we have no evidence of any DIG from irregulars occupying the 
high \nii/\ha part of the diagram. About the first aspect mentioned above 
(low \nii/\ha), given the fact that the position of an emission line object 
in the BPT diagram is related to specific physical conditions of the gas, 
our results suggest that the so-called DIG from irregular galaxies could 
in fact just be the manifestation of low surface brightness \hii regions. 

On the other hand, the lack of DIG from irregulars at high \nii/\ha could 
have several alternative explanations. First, it could be that this emission 
is too weak to have been detected yet. Another possible explanation follows 
from the idea developed in the previous section, namely, that the DIG ionization 
results from ``old stars''. It is conceivable that in irregular galaxies the 
old stellar population is proportionally much smaller than in spiral galaxies, 
the effect of which would be that this population could not then contribute 
significantly to the ionization budget. In any event, since we only have a 
few data points in the case of irregular galaxies (7 \hii regions and only 
2 DIGs with measurements of \nii and \oiii), we cannot yet reach definitive 
conclusions about irregulars.


\begin{table}[!t]
\vspace{0.1cm}
\centering
\setlength{\tabnotewidth}{0.99\columnwidth}
\tablecols{6} 
\caption{Quantil values for \nii/\ha}
\label{quantiles}
\footnotesize 
\begin{tabular}{lccccc}
\toprule
 & {\bf D1} & {\bf Q1} & {\bf Median} & {\bf Q3} & {\bf D9} \\ 
\midrule
\hii reg & $-$0.89 & $-$0.68 & $-$0.52 & $-$0.41 & $-$0.33 \\
DIG      & $-$0.51 & $-$0.38 & $-$0.22 & $-$0.04 &  ~~0.09 \\ 
\bottomrule
\end{tabular}
\vspace{0.2cm}
\end{table}


\section{Conclusions}
\label{sec:conclusions}

In this work, we present the first emission line database from DIGs (\ddb) 
made up from spectroscopic data available in the literature. The database 
is a compilation of 17 bibliographical references. It contains 1061 observed 
regions (309 \hii regions, 218 transition zones and 509 DIGs) out of 29 
galaxies (spirals and irregulars) and is freely available from CDS.

\ddb has allowed us to carry out for the first time a statistical analysis 
that aimed at characterizing the general behavior of the strongest emission 
lines, and at finding global trends among the various line ratios, including 
the EM. One of the results is that the DIGs of irregular galaxies show extreme 
values in their line ratios with respect to that of spirals (see \S~\ref{sec:ID} 
and \S~\ref{sec:discussion}). These, in fact, lie close to the values observed 
in  \hii regions. Since the available data are very limited, it is indispensable 
to obtain more observations to confirm this result.

The analysis carried out with \ddb leads us to define a universal criterion 
for spiral galaxies, which allows distinguishing DIGs from \hii regions. The 
distribution of \nii/\ha ratios shows that \hbox{\nii/\ha $= -0.4$} defines 
a critical value: any emission region with a ratio greater than $-0.3$ should 
be classified as DIG while a ratio below $-0.5$ is likely to come from an \hii region. 

Taking advantage of the high number of data in \ddb, we could confirm (or 
refute) the reality of line ratio behaviors that have been proposed in the 
literature concerning individual galaxies. We could check to what extent 
these applied to all DIGs taken together.  This is the case for the \nii/\ha 
ratio, which shows a well defined anti-correlation with EM, but of varying 
slopes between galaxies. These variations in the log-log slopes rule out 
any definition of a universal EM value to distinguish DIGs from \hii regions. 
Nevertheless, a critical value, EM$_{\rm c}$, can be defined for each galaxy 
individualy.

\vspace{0.15cm}

\acknowledgements

Our thanks for the referee's comments and suggestions that have been very 
relevant for improving and clarifying the manuscript. The authors wish to 
thank Leonid Georgiev who stimulated the writing of this paper. The computations 
were carried out on a AMD-64bit computer financed by grant PAPIIT IX125304 
from DGAPA (Universidad Nacional Aut\'onoma de M\'exico, Mexico). N. F-F. 
is supported by a Conacyt PhD fellowship, C.M. is partly supported by PAPIIT IN123309 from DGAPA (UNAM,Mexico), and Conacyt 
grant 49737, while L.B. is supported by Conacyt grant J-49594.  Some founds also come from grants  PAPIIT IN123309 from DGAPA (UNAM,Mexico).

\end{document}